\documentclass[aps,preprint,amsmath,amssymb]{revtex4}
\usepackage{graphicx,color,ulem}
\def\be{\begin{eqnarray}}
\def\ed{\end{eqnarray}}
\def\non{\nonumber}
\def\ga{\gamma}

\def\la{\langle}
\def\ra{\rangle}
\def\D{\Delta}

\begin{document}

\title{\Large \bf Charge and CP asymmetries of $B_q$ meson\\ in  unparticle physics}

\date{\today}

\author{\bf
Chuan-Hung~Chen$^{1,2}$\footnote{E-mail: physchen@mail.ncku.edu.tw
}, C.~S.~Kim$^{3}$\footnote{E-mail: cskim@yonsei.ac.kr,
}  and Run-Hui Li$^{3}$\footnote{E-mail:lirh@cskim.yonsei.ac.kr}}
\affiliation{
$^{1}$ Department of Physics, National Cheng-Kung University, Tainan 701, Taiwan \\
$^{2}$National Center for Theoretical Sciences, Hsinchu 300, Taiwan
\\
$^{3}$ Department of Physics $\&$ IPAP, Yonsei University, Seoul 120-479, Korea
}

\begin{abstract}
\noindent Recently the D{\O} Collaboration reported an observation
of like-sign charge asymmetry (CA), which is about $3.2~\sigma$
deviation from the standard model (SM) prediction. Inspired by the
observation we investigate the scalar unparticle effects, under the
color charge of $SU(3)_c$ symmetry, in the CP violation in neutral B
meson oscillations as well as the dispersive  and absorptive parts
of $\bar B_q\leftrightarrow B_q$ transition, which can be related to
the CA directly. In order to illustrate the peculiar properties of
unparticle, our analysis is carried out in two scenarios for the
right-handed section: (I) $\lambda_R=\lambda_L$ and $U_D^R=U_D^L$,
where $\lambda_{L,R}$ and $U_D^{L,R}$ are the couplings and flavor
mixing matrix of left- and right-handed section, respectively; (II)
$\lambda_R >> \lambda_L$ and $U_D^R$ is completely a free parameter.
In scenario I we found that the wrong- and like-sign CA cannot be
changed significantly for a SM-like CP violating source because of
the strong constraint of $\Delta m_{B_d}$. Contrarily, in scenario
II we can figure out the parameter space in which the CA can be
enhanced to the value observed by D{\O} with the constraint of
$\Delta m_{B_s}$ due to the enhancement of
$\Gamma^s_{12}$. In the parameter space we obtained, the
correlation between $\Delta\Gamma^s$ and $\phi_s$ is consistent with
the current CDF and D{\O} results.

\end{abstract}

\maketitle

D{\O} Collaboration of Tevatron recently observed the like-sign charge asymmetry (CA), defined as \cite{Abazov:2010hv}
 \be
A^{b}_{s\ell} &=& \frac{N^{++}_b - N^{--}_{b}}{N^{++}_{b}+N^{--}_{b}} \label{eq:Absl_exp}
 \ed
with $N^{++(--)}_b$ being the number of events that $b$-
and $\bar b$-hadrons semileptonically decay into
two positive (negative) muons. The measured value in the dimuon
events is
 \cite{Abazov:2010hv}
 \be
 A^{b}_{s\ell} = \left[-9.57 \pm 2.51({\rm stat}) \pm 1.46 ({\rm syst}) \right]\times 10^{-3}\,.
 \label{eq:d0}
 \ed
Surprisingly, the observation is about 3.2 $\sigma$
away from the SM prediction \cite{Abazov:2010hv,Lenz:2006hd}
of
$$
A^{b}_{s\ell}({\rm SM}) = \left[-2.3^{+0.5}_{-0.6} \right]\times 10^{-4}~.
$$
Since the CA is directly related to CP violation (CPV) in
$B_{d,s}$-meson oscillations and associated with dispersive
($M^q_{12}$) and absorptive ($\Gamma^q_{12}$) parts of $ \bar B_q
\leftrightarrow B_q$ transition,  the large deviations from the SM
could be ascribed  to new CP phases in $b\to d$ and $b\to s$
transitions
\cite{Randall:1998te,Dighe:2010nj,Dobrescu:2010rh,Soni:2010xh,Nandi:2010zx,Chen:2010wv,Buras:2010mh,Ligeti:2010ia,Bauer:2010dg,Deshpande:2010hy,Choudhury:2010ya,Bobrowski:2009ng,Buchalla:2000sk,Altmannshofer:2009ne,Parry:2010ce,Ko:2010mn,King:2010np,Delaunay:2010dw,Bai:2010kf,Kubo:2010mh,Blum:2010mj,Wang:2010vv,Dutta:2010ji,Oh:2010vc,He:2010fz,Park:2010sg,Chao:2010mq,Alok:2010ij,Kim:2010gx,Lee:2010hv,Datta:2010yq}.

Inspired by the anomalous CA, we study the contributions of scale or
conformal invariant stuff, which is known as unparticle
\cite{Georgi1,Georgi2}. The unique character of unparticle is its
peculiar phase appearing in the off-shell propagator with positive
squared transfer momentum \cite{Georgi1,Georgi2,un1,un2}. Due to CP
invariance, the imaginary part of the phase factor leads to the
absorptive effect of a process. In the case of $B_q-\bar B_q$
mixing, not only  can the $M^{q}_{12}$ but also $\Gamma^q_{12}$
be affected \cite{He:2010fz,un2,Lenz:2007nj}. It
is interesting to investigate whether the influence of unparticle on
$M^q_{12}$ and $\Gamma^q_{12}$ could enhance the phase
$\phi_q=arg(-M^q_{12}/\Gamma^q_{12})$ which is directly related to
the CA. In order to make the production of scale invariant stuff be
efficient at Large Hadron Collider (LHC), we investigate the
unparticle that carries the color charges of $SU(3)_c$ symmetry
\cite{Cacciapaglia:2007jq}.

To understand the like-sign CA, we start with discussing the
relevant phenomena.  With strong interaction
eigenbasis, the Hamiltonian for unstable $\bar B_q$ and $B_q$ states
is written as
 \be
 {\bf H }={\bf  M^q }- i \frac{{\bf \Gamma^q }}{2}\,,
 \ed
where ${\bf \Gamma^{q}(M^q)}$ denotes the absorptive (dispersive)
part of the $ \bar B_q \leftrightarrow B_q $ transition.
Accordingly, the time-dependent  wrong-sign CA in semileptonic $B_q$
decays  is defined and given \cite{PDG08} by
 \be
a^q_{s\ell}&=& \frac{\Gamma(\bar B_q(t) \to \ell^+ X)- \Gamma( B_q(t) \to \ell^- X)}{\Gamma(\bar B_q(t) \to \ell^+ X)
+\Gamma( B_q(t) \to \ell^- X)}\,,\non \\
&\approx& Im\left( \frac{\Gamma^{q}_{12}}{M^{q}_{12}}\right) \label{eq:aqsl}\,.
 \ed
Here,  the assumption $\Gamma^q_{12}\ll M^q_{12}$ in
$B_q$ system has been used. Intriguingly, $a^q_{s\ell}$ indeed is
not a time dependent quantity.  The SM predictions
\cite{Lenz:2006hd} are
$$a^d_{s\ell}(\rm SM)=(-4.8^{+1.0}_{-1.2})\times 10^{-4}, ~~~a^{s}_{s\ell}(\rm
SM)=(2.06\pm 0.57)\times 10^{-5},$$ while the
current data \cite{TheHeavyFlavorAveragingGroup:2010qj, Abazov:2009wg} are
$$a^d_{s\ell}(\rm Exp)=(-4.7\pm 4.6)\times 10^{-3},~~~
a^{s}_{s\ell}(\rm Exp)=(-1.7\pm 9.1)\times 10^{-3}.$$

The relation
between the wrong and like-sign CAs is defined and expressed \cite{Abazov:2010hv, Grossman:2006ce}
by
 \be
 A^b_{s\ell} &=& \frac{\Gamma(b\bar b\to \ell^+ \ell^+ X) - \Gamma(b\bar b\to
 \ell^- \ell^- X)}{\Gamma(b\bar b\to \ell^+ \ell^+ X) + \Gamma(b\bar b\to \ell^- \ell^- X)}\,,\non\\
 &=& 0.506(43) a^d_{s\ell} + 0.494(43) a^s_{s\ell}\,. \label{eq:Absl}
 \ed
Clearly, the like-sign CA is associated with the wrong-sign CAs of
$B_d$ and $B_s$ systems. Since the direct measurements of
$a^d_{s\ell}$ and $a^s_{s\ell}$ are still quite uncertain,  either
$b\to  d$ or $b\to s$ transition or both could be the source of
unexpected large $A^b_{s\ell}$.

In order to explore the new physics effects, we  write the
transition matrix elements as $M^q_{12}=M^{q, {\rm SM}}_{12} + M^{q,
{\rm NP}}_{12}$ and $\Gamma^{q}_{12}=\Gamma^{q,{\rm
SM}}_{12}+\Gamma^{q,{\rm NP}}_{12}$ and parameterize them
as
 \be
M^{q}_{12} &=& M^{q,{\rm SM}}_{12} \Delta^M_q e^{i\phi^{\Delta}_q }\,, \non \\
\Gamma^q_{12} &=& \Gamma^{q, {\rm SM}}_{12} \Delta^{\Gamma}_{q} e^{i\ga^{\Delta}_q} \label{eq:MG}
\ed
with
 \be
 M^{q, {\rm SM}[{\rm NP}]}_{12}&=& \left|M^{q, {\rm SM}[{\rm NP}]}_{12}\right| e^{2i\bar\beta_q[\theta^{{\rm NP}}_{q}]}\,,\ \   \Gamma^{q, {\rm SM}}_{12}=\left |\Gamma^{q, {\rm SM}[{\rm NP}]}_{12}\right| e^{i\gamma^{{\rm SM}[{\rm NP}]}_q}\,, \non \\
 \Delta^M_q  &=& \left| 1 + r^M_q e^{2 i (\theta^{{\rm NP}}_q -\bar\beta_q)}\right|\,, \ \  r^M_q = \frac{|M^{q,{\rm NP}}_{12}|}{|M^{q,{\rm SM}}_{12}|}\,, \non \\
\Delta^\Gamma_q  &=& \left| 1 + r^\Gamma_q e^{i(\ga^{{\rm NP}}_q -\ga^{\rm SM}_q)}\right|\,, \ \  r^\Gamma_q = \frac{|\Gamma^{q,{\rm NP}}_{12}|}{|\Gamma^{q,{\rm SM}}_{12}|}\,, \non \\
\tan\phi^\Delta_q &=& \frac{r^M_q \sin2(\theta^{{\rm NP}}_{q} -\bar\beta_q)}{1 + r^M_q \cos2(\theta^{{\rm NP}}_{q} -\bar\beta_q)}\,, \ \
\tan\ga^\Delta_q = \frac{ r^\Gamma_q \sin(\ga^{{\rm NP}}_{q} -\ga^{\rm SM}_{q})}{1 - r^\Gamma_q \cos(\ga^{{\rm NP}}_{q} -\ga^{\rm SM}_{q})}\,.
  \ed
The phases appearing above stand for weak CP violating phases. We
note that although $\bar\beta_q$ is not a conventional
notation for the CP phase of the SM denoted by $\beta_q$, their
relationship could be read by $\bar\beta_d=\beta_d$ and
$\bar\beta_s=-\beta_s$. Using $\phi_q =
arg(-M^q_{12}/\Gamma^q_{12})$, the wrong-sign CA in
Eq.~(\ref{eq:aqsl}) with new physics effects on $\Gamma^q_{12}$ and
$M^q_{12}$ could be given as
 \be
 a^q_{s\ell} &=&\frac{\Delta^\Gamma_q}{\Delta^M_q} \frac{\sin\phi_q}{\sin\phi^{\rm SM}_{q}}a^{q}_{s\ell}({\rm SM}) \label{eq:aqsl_new}
 \ed
with $\phi^{\rm SM}_q=2\bar\beta_q -\ga^{\rm SM}_{q}$ and $\phi_q=\phi^{\rm SM}_{q} + \phi^{\Delta}_{q}-\ga^{\D}_{q}$.
Furthermore, the mass and rate differences between
heavy and light $B$ mesons could be expressed by
 \be
 \Delta m_{B_q}&=& 2|M^q_{12}|\,,\non\\
 \D\Gamma^q &=&\Gamma_L-\Gamma_H= 2 |\Gamma^q_{12}| \cos\phi_q\,. \label{eq:delga}
 \ed

Another type of  the time-dependent CP asymmetry (CPA) is associated
with the definite CP  in the final state, defined  by \cite{PDG08}
 \be
A_{f_{CP}}(t)&\equiv& \frac{\Gamma(\bar B_q(t) \to f_{CP})- \Gamma( B_q(t) \to f_{CP})}{\Gamma(\bar B_q(t) \to f_{CP})
+\Gamma( B_q(t) \to f_{CP})}\,,\non \\
&=& S_{f_{CP}} \sin\Delta m_{B_q} t - C_{f_{CP}} \cos\Delta m_{B_q} t\,, \non\\
S_{f_{CP}} &=& \frac{2Im\lambda_{f_{CP}} }{1+|\lambda_{f_{CP}}|^2}\,,\ \ \ C_{f_{CP}}
= \frac{1-|\lambda_{f_{CP}}|^2}{1+|\lambda_{f_{CP}}|^2}
\label{eq:Sf}
 \ed
with
 \be
 \lambda_{f_{CP}} &=&\left(\frac{M^{B_q^*}_{12}}{M^{B_q}_{12}}\right)^{1/2}
  \frac{A(\bar B\to f_{CP})}{A(B\to f_{CP})} \,, \label{eq:lambdaf}
 \ed
where $f_{CP}$ denotes the final CP eigenstate, $S_{f_{CP}}$ and
$C_{f_{CP}}$ are the so-called mixing-induced and direct
CPAs, respectively. Clearly, beside the phases  in the
$\Delta B=2$ processes, the mixing-induced CPA is also
related to the phase in the $\Delta B=1$ process. Nevertheless,
since the new effects on the decays $B_d\to J/\Psi K_S$ and  $B_s\to
J/\Psi \phi$ are small, the CPAs could be simplified as
 \be
 S_{J/\Psi K_S}&\equiv& \sin2\bar\beta_{J/\Psi K_S} \approx  \sin(2\bar\beta_d + \phi^{\Delta}_{d})\,, \non \\
 S_{J/\Psi \phi} &\equiv& \sin2\bar\beta^{J/\Psi \phi}_{s} \approx \sin(2\bar\beta_s + \phi^{\Delta}_{s})\,. \label{eq:Sjpsi_phi}
 \ed

After introducing the relevant physical observables, we begin
studying the effects of colored scalar unparticle.  Since there is
no well established approach to give a full theory for unparticle
interactions, we study the topic from the phenomenological
viewpoint. In order to avoid fine-tuning the parameters for  flavor
changing neutral currents (FCNCs) at tree level, we assume that the
unparticle only couples to the third generation of quarks before
electroweak symmetry breaking. Hence, the interactions obeying the
SM gauge symmetry are expressed by
 \be
 \frac{1}{\Lambda^{d_U}_U}\left[ \lambda_R \bar q'_R \ga_\mu  T^a q'_R \partial^\mu {\cal O}^{a}_U +
 \lambda_L \bar Q_L \ga_\mu  T^a Q_L \partial^\mu {\cal O}^a_U\right]\,,
\label{eq:lag_un}
 \ed
where $\lambda_{R, L}$ are dimensionless free parameters,
$q'_R=t_R$, $b_R$, $Q^T_L=(t, b)_L$, $\{T^a\} = \{\lambda^a/2\}$ are
the $SU(3)_c$ generators (where $\lambda^a$ are the Gell-Mann
matrices) normalized by $tr (T^a T^b) = \delta^{ab}/2$, $\Lambda_U$
is the scale below which the unparticle is formed. The power $d_U$
is determined from the effective interaction of
Eq.~(\ref{eq:lag_un}) in four-dimensional space-time when the
dimension of the colored unparticle ${\cal O}^{a}_U$ is taken as
$d_U$. Since we only concentrate on the phenomena of down type
quarks, the associated pieces are formulated by
 \be
 \bar D \ga_\mu \left( {\bf X}_R P_R + {\bf X}_L P_L\right) T^a D \partial^\mu {\cal O}^a_U\,,
 \label{eq:int_un}
 \ed
in which  $D^T=(d, s, b)$, ${\bf X}_{R(L)}$ is a $3\times 3 $
diagonal matrix and diag(${\bf X}_{R(L)}$)=(0, 0,
$\lambda_{R(L)}/\Lambda^{d_U}_{U}$). After spontaneous symmetry
breaking of electroweak symmetry, we need to introduce two unitary
matrices $U^{R, L}_D$ to diagonalize the mass matrix of down type
quarks. In terms of physical eigenstates and using the equations of
motion, the interactions for $b-q-{\cal O}^a_U$ could be written as
 \be
 {\cal L}_{bq{\cal O}^a_U}&=& \frac{m_b}{\Lambda^{d_U}_{U}} \bar q \left( f^R_{qb} P_L + f^L_{qb} P_R\right)
 T^a b{\cal O}^a_U+h.c.\,, \label{eq:int_bq}
 \ed
where $q=d, s$, the mass of light quark has been neglected and
$f^{\chi}_{qb} = \lambda_{\chi}(U^{\chi}_D)_{qb}
(U^{\chi^*}_D)_{bb}$ with $\chi=R, L$.

By following the scheme shown in Ref.~\cite{Grinstein:2008qk}, the propagator of the colored scalar unparticle is written as
 \be
\int d^4x e^{-ik\cdot x} \la 0|T {\cal O} ^{a}(x) {\cal
O}^{b}(0)|0\ra  = i \frac{C_S \delta^{ab}}{(-k^2
-i\epsilon)^{2-d_U}} \label{eq:prop_un}
 \ed
with
 \be
 C_S &=& \frac{A_{d_U}}{2\sin d_{U} \pi} \,, \non\\
 A_{d_U}&=& \frac{16\pi^{5/2}}{(2\pi)^{2d_U}} \frac{\Gamma(d_U +1/2)}{\Gamma(d_U-1) \Gamma(2d_U)}\,.
 \ed
Combining Eqs.~(\ref{eq:int_bq}) and (\ref{eq:prop_un}), the four
fermion interaction for $B_q$ oscillation is given by \be {\cal H}
&=& \frac{C_S }{2m^2_b }
\left(\frac{m^2_b}{\Lambda^2_U}\right)^{d_U} f^2_{qb} e^{-id_U \pi}
\left( \bar q T^a b \right)^2\,. \ed For estimating the
transition matrix elements,  we employ the vacuum insertion method
and the results are
 \be
 \la \bar B_q | \bar q P_{R(L)} b \bar q P_{R(L)} b| B_q \ra &\approx&  -\frac{5}{24} m_{B_q} f^2_{B_q} \,, \non \\
 \la \bar B_q | \bar q P_{R} b \bar q P_{L} b| B_q \ra &\approx&   \frac{7}{24} m_{B_q} f^2_{B_q}\,, \non \\
 \la \bar B_q | \bar q_\alpha P_{R(L)} b_\beta \bar q_\beta P_{R(L)} b_\alpha | B_q \ra &\approx& \frac{1}{24}  m_{B_q} f^2_{B_q}\,, \non \\
 \la \bar B_q | \bar q_\alpha P_{R} b_\beta \bar q_\beta P_{L} b_\alpha | B_q \ra &\approx&  \frac{5}{24} m_{B_q} f^2_{B_q}
 \ed
where the approximation $m_b \sim m_{B_q}$ is used and $f_{B_q}$ is
the decay constant of $B_q$ meson. As a consequence,  the
dispersive and absorptive parts of $\bar B_q\leftrightarrow B_q$ in
the unparticle physics are found by
 \be
H^{U}_{12} &=& M^{q, U}_{12} - i \frac{\Gamma^{q, U}_{12}}{2} \,, \non \\
{\rm where}~~~~M^{q, U}_{12}&=& \cos(d_U\pi) h^q_U\,, \ \ \Gamma^{q, U}_{12}=2\sin(d_U \pi) h^q_U\,,
 \ed
with
 \be
 h^q_U=  \frac{C_S}{18}  \left( f^R_{qb} +f^L_{qb}\right)^2 \left(\frac{m^2_{B_q}}{\Lambda^2_U}\right)^{d_U}  \frac{f^2_{B_q}}{m_{B_q}}\,.
 \ed
For comparison, we also summarize the formulae of the
SM as follows \cite{PDG08}:
 \be
M^{q,SM}_{12}&=&\frac{G^{2}_{F} m^2_{W}}{12 \pi^2}
\eta_{B} m_{B_q}f^{2}_{B_q}\hat{B}_q ( V^*_{tq}V_{tb})^2 S_{0}(x_t)\,, \non \\
\Gamma^{q,SM}_{12}&\approx & \frac{3\pi}{2}\left( \frac{m^2_b}{m^2_W}\right) \frac{M^{q,SM}_{12}}{S_0(x_t)}\left [ 1+ \frac{V^*_{cq} V_{cb}}{V^{*}_{tq} V_{tb}}O(m^2_c/m^2_b) \right]
 \ed
with $S_{0}(x_t)=0.784 x_t^{0.76}$, $x_{t}=(m_t/m_W)^2$ and
$\eta_{B}\approx 0.55$ is the QCD correction to $S_0(x_t)$.

In the considering model, in addition to  the scale dimension $d_U$,
the couplings $\lambda_{R, L}$ and the scale $\Lambda_U$  that are
associated with unparticle,  the flavor mixing elements
$(U^{\chi}_{D})_{qb}(U^{\chi^*}_{D})_{bb}$ in $f^{\chi}_{qb}$ are
also free parameters. Following the Cabibbo-Kobayashi-Maskawa (CKM)
matrix defined by $V=U^L_U U^{L^\dagger}_{D}$, indeed
$(U^{L}_{D})_{qb} = V^*_{tq}$ when we choose the convention
$U^L_U=1$.  If we take the CKM matrix as inputs, then the
right-handed flavor mixing element $(U^R_D)_{qb}$ is the only free
parameter. Therefore, to illustrate the peculiar properties of
unparticle, we study two scenarios for $\lambda_{R,L}$ and $U^R_D$:
(I) $\lambda_R=\lambda_L=\lambda_U$ and $U^R_D = U^L_D=V^\dagger$
(i.e. $f^R_{qb}=f^L_{qb}$); (II) $\lambda_L\ll \lambda_R$ and
$U^R_D$ is unknown (i.e. $f^L_{qb} \ll f^R_{qb}$). In scenario I,
the couplings of unparticle to fermions are vector-like. In scenario
II, since the behavior of left-handed couplings is similar to the
scenario I, for illustrating the influence of right-handed couplings
we set $\lambda_L \ll \lambda_R$. For simplicity, in the numerical
estimates we take $\Lambda_U=1$ TeV.

For numerical calculations and constraints, we list the useful
values in Table~\ref{tab:inputs},  where  the relevant CKM matrix
element $V_{tq}=\bar V_{tq}\exp(-i\bar\beta_q)$ is obtained from the
UTfit Collaboration~\cite{Bona:2009tn}, the decay constant of $B_q$
is referred to the result given  by the HPQCD
Collaboration~\cite{Gamiz:2009ku} and the value of $\phi^{\rm SM}_q$ is
from Ref.~\cite{Lenz:2006hd}. The CDF and D$\O$ average values of
$\Delta \Gamma^s=[-0.163, 0.163]$ and  $\phi_s= [-1.35 , -0.20 ]
\cup [ -2.94 ,  -1.77 ]$ with $90\%$ confidence level (CL) are from
Ref.~\cite{TheHeavyFlavorAveragingGroup:2010qj}.  Other inputs are
quoted from the particle data group (PDG) \cite{PDG08}. As a result,
we obtain $|M^{d, {\rm SM}}_{12}|=0.253$ ps$^{-1}$, $|M^{s,
{\rm SM}}_{12}|=8.90$ ps$^{-1}$. In addition, according to the results in
Ref.~\cite{Lenz:2006hd}, we also know $\Gamma^{d, {\rm SM}}_{12}\approx
-1.3 \times 10^{-3} \exp[i(2\beta_d -\phi^{\rm SM}_d)]$ ps$^{-1}$ and
$\Gamma^{s, {\rm SM}}_{12}\approx  -0.048 \exp[i(2\beta_s-\phi^{\rm SM}_{s})]$
ps$^{-1}$.
\begin{table}[tb]
\caption{Experimental data and numerical inputs for the parameters in the SM.
 } \label{tab:inputs}
\begin{ruledtabular}
\begin{tabular}{cccc}
  $\bar V_{td}$ & $\bar\beta_d$ & $ \bar V_{ts}$ & $\bar\beta_s$
 \\ \hline
 $8.51(22)\times 10^{-3}$ & $0.384\pm 0.014$ & $ -4.07(22)\times 10^{-2}
 $ & $-0.018\pm 0.001$  \\ \hline\hline
  $f_{B_d} \sqrt{\hat B}_d$ & $f_{B_s} \sqrt{\hat{B_s}}$ & $f_{B_d}$ & $f_{B_s}$
   \\ \hline
 $(216\pm 15)$ MeV & $(266\pm 18)$ MeV & $190\pm 13 $ MeV & $231 \pm 15$ MeV   \\ \hline\hline
  $(\Delta m_{B_d})^{\rm Exp}$ & $(\Delta m_{B_s})^{\rm Exp}$  &  $\phi^{SM}_{d}$ &  $\phi^{SM}_s$   \\ \hline
 $0.507 \pm 0.005$ ps$^{-1}$ & $17.77 \pm 0.12$ ps$^{-1}$ &  $-0.091^{+0.026}_{-0.038}$  &
  $(4.2\pm 1.4)\times 10^{-3}$ \\

\end{tabular}
\end{ruledtabular}
\end{table}

\begin{figure}[tb]
\includegraphics*[width=5.0 in]{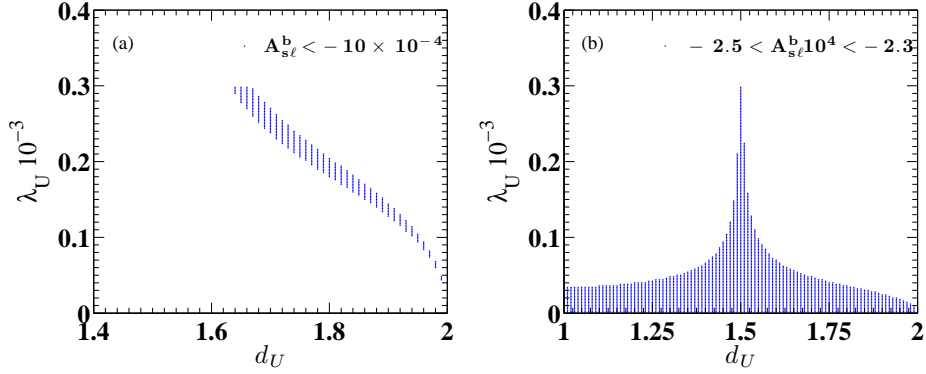}
\caption{ The allowed range of $\lambda_U$ and $d_U$ constrained by
(a) $S_{J/\Psi K_S}$ and (b) $\Delta m_{B_d}$, where the available
range of $A^b_{s\ell}$  is  (a) less than $-10 \times 10^{-4}$ and
(b) within $(-2.5, -2.3)\times 10^{-4}$.}
 \label{fig:con_d}
\end{figure}

We first discuss the situation in scenario I, i.e. the case with
$f^R_{qb}=f^L_{qb}$. Due to $U^R_{qb} = U^L_{qb}=V^*_{tq}$, the CP
phase for $b\to q$ transition in unparticle exchange is the same as
that in the SM. Therefore, the influence of unparticle on CPAs of
$b\to s$ transition is small and insignificant. Because  $\lambda_U$
and $d_U$ are only the free parameters, it is interesting to see if
the unparticle could have a large effect on the wrong-sign CA. At
first, we only consider the constraint from the time-dependent CPA
of $B_d$ which is formulated in Eqs.~(\ref{eq:Sf}) and
(\ref{eq:Sjpsi_phi}) and measured with $S_{J/\Psi K_S}=0.655\pm
0.0244$ \cite{TheHeavyFlavorAveragingGroup:2010qj}.  Taking the data
of $S_{J/\Psi K_S}$ with $2\sigma$ errors as the constraint, we find
that $A^b_{s\ell} < -10\times 10^{-4}$ could be archived. The
allowed region for $\lambda_U$ and $d_U$ is shown in
Fig.~\ref{fig:con_d}(a). Unfortunately, the enhancement on the
magnitude of $A^b_{s\ell}$ is suppressed when we include the
constraint from the measurement of $\Delta m_{B_d}$. Taking the
$(\Delta m_{B_d})^{\rm Exp}$ with $2\sigma$ errors as the
constraint, we find that the resulted like-sign CA is close to the
SM prediction. The allowed region of the parameters constrained by
$\Delta m_{B_d}$ are presented in Fig.~\ref{fig:con_d}(b), where the
available range for like-sign CA is $-2.5<A^b_{s\ell}10^{4}<-2.3$ .
We see clearly that if the CP violating source is SM-like, by the
strong constraint of $\Delta m_{B_d}$, the wrong- and like-sign CA
cannot be changed significantly.

\begin{figure}[t]
\includegraphics*[width=5.5 in]{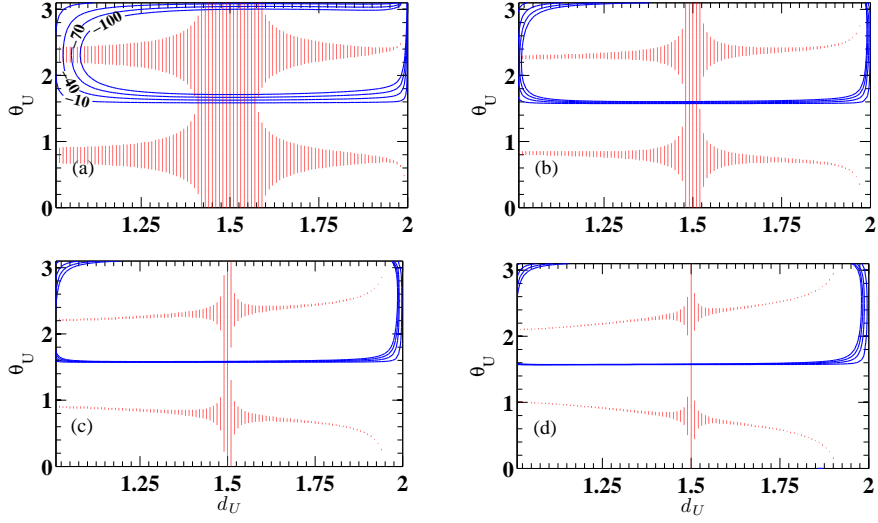}
\caption{(a)-(d) bound of $\Delta m_{B_s}$ (scatters) and $-100\leq
A^{b}_{s\ell}10^{4}\leq -10$ (solid) for $|f^R_{sb}|=(4, 8, 12,
16)\times 10^{-6}$, respectively.   }
 \label{fig:con_s}
\end{figure}

Next, we study the phenomena in scenario II. As stated early, the
effects of left-handed coupling are similar to the case of scenario
I, in order to display the peculiar property of unparticle, we set
$\lambda_L\ll \lambda_R$ so that $f^L_{qb} \ll f^R_{qb}$.
Additionally, since $\Delta m_{B_d}$ and $S_{J/\Psi K_S}$ will give
a strong constraint on the parameters for $b\to d$ transition
\cite{Chen:2010wv}, here we only concentrate on the phenomena
associated with  $b\to s$ transition. Due to $\lambda_R$ and $U^R_D$
being unknown, we use complex $f^R_{sb}$ as the variable. In order
to simplify the analysis, we will choose some specific values for
$|f^R_{sb}|$ and vary the phase $\theta_U=arg(f^R_{sb})$ within $[0,
\pi]$. The results of $[-\pi, 0]$ are expected to be similar to
those in $[0, \pi]$. Consequently, with $2\sigma$ errors of $(\Delta
m_{B_s})^{\rm Exp}$, we display the constraint on $\theta_U$ and
$d_U$ in Fig.~\ref{fig:con_s}, where the figure (a)-(d) respectively
corresponds to $|f^R_{sb}|=(4, 8, 12, 16)\times 10^{-6}$ and the
scatters represent the bound given by $\Delta m_{B_s}$.  In terms of
Eqs. (\ref{eq:Absl}) and (\ref{eq:aqsl_new}),  we plot $-100\leq
A^b_{s\ell}10^{4}\leq -10$ which is induced by the unparticle in
Fig.~\ref{fig:con_s}. We find that with $|f^R_{sb}|=4\times
10^{-6}$, the enhanced like-sign CA could occur at $1<d_U <1.1$ with
$2.2<\theta_U <2.4$ and a wider region around $d_U\sim 3/2$ with
$\theta_U\sim 1.6$. For $|f^R_{sb}|=8\times 10^{-6}$, only $d_U\sim
1$ with $\theta_U\sim 2.2$ and $d_U\sim 3/2$ with $\theta_U\sim 1.6$
can have large $-A^b_{s\ell}$.  As to other values of $|f^R_{sb}|$,
they only happen at $d_U\sim 3/2$ and $\theta_U\sim 1.6$.
\begin{figure}[b]
\includegraphics*[width=5.5 in]{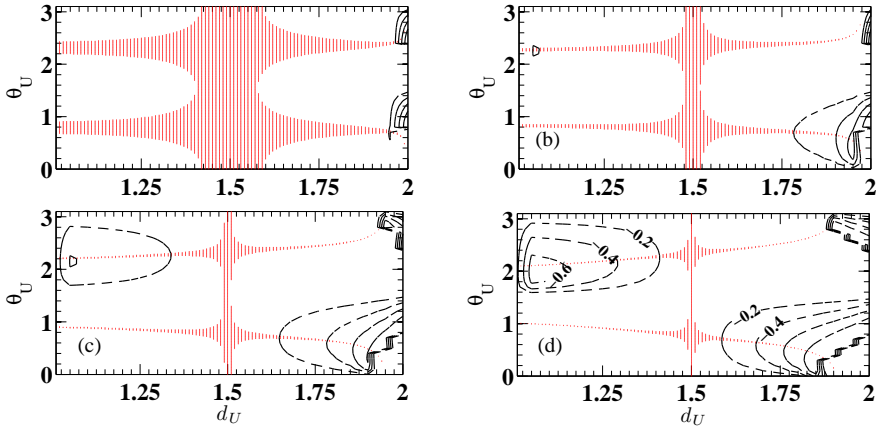}
\caption{The Legend is the same as that in Fig.~\ref{fig:con_s} but for $-1<S_{J/\Psi \phi}<-0.2$ (dashed).   }
 \label{fig:S_Jpsiphi}
\end{figure}

Similarly, we can use the same approach to study the influence of
unparticle on the time-dependent CPA of Eq.~(\ref{eq:Sjpsi_phi}).
Taking the data $\phi_s= [-1.35 , -0.20 ] \cup [ -2.94 ,  -1.77 ]$,
i.e.  $-1<S_{J/\Psi \phi} <-0.2$
\cite{TheHeavyFlavorAveragingGroup:2010qj}, we display the contour
of $S_{J/\Psi \phi}$ as a function of $d_U$ and $\theta_U$ in
Fig.~\ref{fig:S_Jpsiphi}, where  figure (a)-(d) corresponds to
$|f^R_{sb}|=(4,8,12,16)\times 10^{-6}$ respectively, the scatters
stand for the constraint of $\Delta m_{B_s}$ and the dashed lines
denote $-1<S_{J/\Psi \phi} <-0.2$. By comparing
Fig.~\ref{fig:S_Jpsiphi} with Fig.~\ref{fig:con_s}, we find that
large $-A^b_{s\ell}$ and $-S_{J/\Psi \phi}$ induced by the
unparticle exchange cannot exist simultaneously. It is interesting if the
peculiar results could be confirmed in the Super B factories,
Tevatron and LHCb. Then, we would have more strong evidence to
believe the existence of scale invariant stuff.

\begin{figure}[tb]
\includegraphics*[width=5.5 in]{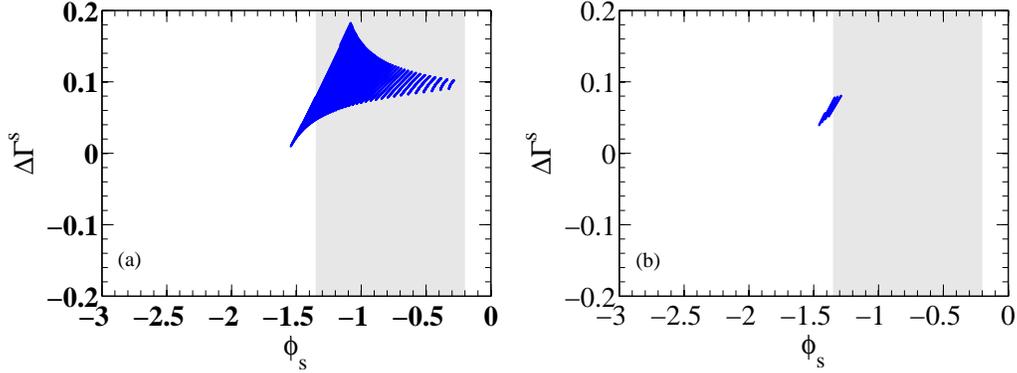}
\caption{ (a)[(b)] Correlation of $\Delta\Gamma^s$ with $\phi_s$ for
$f^R_{sb}=4[8]\times 10^{-6}$, where the constraint of $(\Delta
m_{B_s})^{\rm Exp}$  with $2\sigma$ errors is included and $-100<
A^b_{s\ell}10^{4}<-10$ has been archived. }
 \label{fig:delga_phis}
\end{figure}

Beside the like-sign CA, $A^b_{s\ell}$, and time-dependent CPA,
$S_{J/\Psi \phi}$, it is also important to study the correlation of
$\Delta\Gamma^s$ with $\phi_s$  defined in Eq.~(\ref{eq:delga}).
Thus, in terms of Eqs.~(\ref{eq:MG}), (\ref{eq:delga}) and the
definition $\phi_s=arg(-M^s_{12}/\Gamma^s_{12})$, the correlation
between $\Delta\Gamma^s$ and $\phi_s$ resulted by the allowed values
of $d_U$ and $\theta_U$ that satisfy $(\Delta m_{B_s})^{\rm Exp}$
with $2\sigma$ errors and $-100<A^b_{s\ell}10^{4}<-10$ is presented
in Fig.~\ref{fig:delga_phis}, in which the bands in the figure
denote the data. We see that only the cases of $|f^R_{sb}|=(4, 8)
\times 10^{-6}$ can be consistent with the current data of
$\Delta\Gamma^s$ and $\phi_s$ when the bound of $\Delta m_{B_s}$ is
included and large $-A^b_{s\ell}$ is archived. By the figure, we
learn that the smaller $|f^R_{sb}|$ owns a wider range of $\phi_s$.
This behavior could be understood from Fig.~\ref{fig:con_s}(a) and
(b) where the available $d_U$ in the former is much wider than that
in the latter.

In summary, we have studied the peculiar phase of unparticle on
$M^q_{12}$ and $\Gamma^{q}_{12}$. In order to produce the unparticle
efficiently at the LHC, we investigated the colored unparticle on
the like-sign CA and time-dependent CPA with two scenarios of the
free parameters chosen. In the scenario I{\color{red} ,} in which
the involved CP phase is the same as that in the SM, the like-sign
CA could be enhanced largely with the constraint of $S_{J/\Psi K_S}$
only. However, the CA becomes suppressed when the constraint of
$\Delta m_{B_d}$ is taken into account.  In the scenario II where
the new CP phase is from the right-handed flavor mixing matrix, we
find that $A^b_{s\ell}$ could be enhanced to the value observed by
D{\O}, whereas the corresponding time-dependent CP cannot be
enhanced to the range of current data. Additionally, the correlation
between $\Delta\Gamma^s$ and $\phi_s$ could be consistent with
current CDF and D{\O} results while the constraint of $\Delta
m_{B_s}$ is taken into account and $-100<A^b_{s\ell}10^{4}<-10$ is
archived.

\begin{acknowledgments}
\noindent C.H.C was supported in part by the National Science
Council of R.O.C. under Grant No. NSC-97-2112-M-006-001-MY3. The
work of C.S.K.  was supported in part by the Basic Science Research
Program through the NRF of Korea funded by MOEST (2009-0088395) and
in part by KOSEF through the Joint Research Program
(F01-2009-000-10031-0). The work of R.H.L. was supported by Brain
Korea 21 Project.
\end{acknowledgments}



\begin{thebibliography}{99}

\bibitem{Abazov:2010hv}
  V.~M.~Abazov {\it et al.}  [D0 Collaboration],
  Phys.\ Rev.\  D {\bf 82}, 032001 (2010)
  [arXiv:1005.2757 [hep-ex]].

\bibitem{Lenz:2006hd}
  A.~Lenz and U.~Nierste,
  JHEP {\bf 0706}, 072 (2007)
  [arXiv:hep-ph/0612167];
  A.~J.~Lenz,
  AIP Conf.\ Proc.\  {\bf 1026}, 36 (2008)
  [arXiv:0802.0977 [hep-ph]];
  A.~S.~Dighe, T.~Hurth, C.~S.~Kim and T.~Yoshikawa,
  Nucl.\ Phys.\  B {\bf 624} (2002) 377
  [arXiv:hep-ph/0109088].


\bibitem{Randall:1998te}
 L.~Randall and S.~f.~Su,
 Nucl.\ Phys.\  B {\bf 540}, 37 (1999)
 [arXiv:hep-ph/9807377].

\bibitem{Dighe:2010nj}
   A.~Dighe, A.~Kundu and S.~Nandi,
  Phys.\ Rev.\  D {\bf 82}, 031502 (2010)
  [arXiv:1005.4051 [hep-ph]].

\bibitem{Dobrescu:2010rh}
  B.~A.~Dobrescu, P.~J.~Fox and A.~Martin,
  Phys.\ Rev.\ Lett.\  {\bf 105}, 041801 (2010)
  [arXiv:1005.4238 [hep-ph]].

\bibitem{Soni:2010xh}
  A.~Soni, A.~K.~Alok, A.~Giri, R.~Mohanta and S.~Nandi,
  Phys.\ Rev.\  D {\bf 82}, 033009 (2010)
  [arXiv:1002.0595 [hep-ph]].

\bibitem{Nandi:2010zx}
  S.~Nandi and A.~Soni,
  arXiv:1011.6091 [hep-ph].

\bibitem{Chen:2010wv}
  C.~H.~Chen and G.~Faisel,
  arXiv:1005.4582 [hep-ph];
  C.~H.~Chen, C.~Q.~Geng and W.~Wang,
  arXiv:1006.5216 [hep-ph].

\bibitem{Buras:2010mh}
  A.~J.~Buras, M.~V.~Carlucci, S.~Gori and G.~Isidori,
  arXiv:1005.5310 [hep-ph].

\bibitem{Ligeti:2010ia}
  Z.~Ligeti, M.~Papucci, G.~Perez and J.~Zupan,
  Phys.\ Rev. Lett.\ {\bf 105}, 131601 (2010) [arXiv:1006.0432 [hep-ph]].

\bibitem{Bauer:2010dg}
  C.~W.~Bauer and N.~D.~Dunn,
  arXiv:1006.1629 [hep-ph].


\bibitem{Deshpande:2010hy}
  N.~G.~Deshpande, X.~G.~He and G.~Valencia,
  Phys.\ Rev.\ D {\bf 82}, 056013 (2010) [arXiv:1006.1682 [hep-ph]].



\bibitem{Choudhury:2010ya}
  D.~Choudhury and D.~K.~Ghosh,
  arXiv:1006.2171 [hep-ph].


\bibitem{Bobrowski:2009ng}
  M.~Bobrowski, A.~Lenz, J.~Riedl and J.~Rohrwild,
  Phys.\ Rev.\  D {\bf 79}, 113006 (2009)
  [arXiv:0902.4883 [hep-ph]];
  O.~Eberhardt, A.~Lenz and J.~Rohrwild,
  arXiv:1005.3505 [hep-ph].

\bibitem{Buchalla:2000sk}
  G.~Buchalla, G.~Hiller and G.~Isidori,
  Phys.\ Rev.\  D {\bf 63}, 014015 (2000)
  [arXiv:hep-ph/0006136];
  C.~Bobeth, G.~Hiller and G.~Piranishvili,
  JHEP {\bf 0807}, 106 (2008)
  [arXiv:0805.2525 [hep-ph]];
  C.~Bobeth, G.~Hiller and D.~van Dyk,
  JHEP {\bf 1007}, 098 (2010)
  [arXiv:1006.5013 [hep-ph]].

\bibitem{Altmannshofer:2009ne}
  W.~Altmannshofer, A.~J.~Buras, S.~Gori, P.~Paradisi and D.~M.~Straub,
  Nucl.\ Phys.\  B {\bf 830}, 17 (2010)
  [arXiv:0909.1333 [hep-ph]].



\bibitem{Parry:2010ce}
  J.~K.~Parry,
  arXiv:1006.5331 [hep-ph].

\bibitem{Ko:2010mn}
  P.~Ko and J.~h.~Park,
  arXiv:1006.5821 [hep-ph].

\bibitem{King:2010np}
  S.~F.~King,
  JHEP {\bf 1009}, 114 (2010)
  [arXiv:1006.5895 [hep-ph]].

\bibitem{Delaunay:2010dw}
  C.~Delaunay, O.~Gedalia, S.~J.~Lee, G.~Perez and E.~Ponton,
  arXiv:1007.0243 [hep-ph].

\bibitem{Bai:2010kf}
  Y.~Bai and A.~E.~Nelson,
  arXiv:1007.0596 [hep-ph].

\bibitem{Kubo:2010mh}
  J.~Kubo and A.~Lenz,
  Phys.\ Rev.\  D {\bf 82}, 075001 (2010)
  [arXiv:1007.0680 [hep-ph]].

\bibitem{Blum:2010mj}
  K.~Blum, Y.~Hochberg and Y.~Nir,
  JHEP {\bf 1009}, 035 (2010)
  [arXiv:1007.1872 [hep-ph]].

\bibitem{Wang:2010vv}
  R.~M.~Wang, Y.~G.~Xu, M.~L.~Liu and B.~Z.~Li,
  arXiv:1007.2944 [hep-ph].

\bibitem{Dutta:2010ji}
  B.~Dutta, Y.~Mimura and Y.~Santoso,
  Phys.\ Rev.\  D {\bf 82}, 055017 (2010)
  [arXiv:1007.3696 [hep-ph]].

\bibitem{Oh:2010vc}
  S.~Oh and J.~Tandean,
  arXiv:1008.2153 [hep-ph].

\bibitem{He:2010fz}
  X.~G.~He, B.~Ren and P.~C.~Xie,
  arXiv:1009.3398 [hep-ph].


\bibitem{Park:2010sg}
  S.~C.~Park, J.~Shu, K.~Wang and T.~T.~Yanagida,
  arXiv:1008.4445 [hep-ph].

\bibitem{Chao:2010mq}
  W.~Chao and Y.~c.~Zhang,
  arXiv:1008.5277 [hep-ph].


\bibitem{Alok:2010ij}
  A.~K.~Alok, S.~Baek and D.~London,
  arXiv:1010.1333 [hep-ph].

\bibitem{Kim:2010gx}
  J.~E.~Kim, M.~S.~Seo and S.~Shin,
  arXiv:1010.5123 [hep-ph].

\bibitem{Lee:2010hv}
  J.~P.~Lee and K.~Y.~Lee,
  arXiv:1010.6132 [hep-ph].

\bibitem{Datta:2010yq}
  A.~Datta, M.~Duraisamy and S.~Khalil,
  arXiv:1011.5979 [hep-ph].

\bibitem{Georgi1}
  H.~Georgi,
  Phys.\ Rev.\ Lett.\  {\bf 98}, 221601 (2007)
  [arXiv:hep-ph/0703260].

\bibitem{Georgi2}
H.~Georgi,
  Phys.\ Lett.\  B {\bf 650}, 275 (2007)
  [arXiv:0704.2457 [hep-ph]].

\bibitem{un1}
 K.~Cheung, W.~Y.~Keung and T.~C.~Yuan,
  Phys.\ Rev.\ Lett.\  {\bf 99}, 051803 (2007)
  [arXiv:0704.2588 [hep-ph]].

\bibitem{un2}
  C.~H.~Chen and C.~Q.~Geng,
  Phys.\ Rev.\  D {\bf 76}, 115003 (2007)
  [arXiv:0705.0689 [hep-ph]];
{\it ibid} {\bf 76}, 036007 (2007)
  [arXiv:0706.0850 [hep-ph]];
  Phys.\ Lett.\  B {\bf 661}, 118 (2008)
  [arXiv:0709.0235 [hep-ph]];
  C.~H.~Chen, C.~S.~Kim and Y.~W.~Yoon,
  Phys.\ Lett.\  B {\bf 671}, 250 (2009)
  [arXiv:0801.0895 [hep-ph]];
  C.~H.~Chen and C.~S.~Kim,
  Phys.\ Lett.\  B {\bf 687}, 232 (2010)
  [arXiv:0909.1878 [hep-ph]];
 C.~H.~Chen, G.~Cvetic and C.~S.~Kim,
  arXiv:1009.4165 [hep-ph].

\bibitem{Lenz:2007nj}
  A.~Lenz,
  Phys.\ Rev.\  D {\bf 76}, 065006 (2007)
  [arXiv:0707.1535 [hep-ph]].


\bibitem{Cacciapaglia:2007jq}
  G.~Cacciapaglia, G.~Marandella and J.~Terning,
  JHEP {\bf 0801}, 070 (2008)
  [arXiv:0708.0005 [hep-ph]].


\bibitem{PDG08} C. Amsler {\it et al.} [Particle Data Group], Phys. Lett. B{\bf 667}, 1 (2008).

\bibitem{TheHeavyFlavorAveragingGroup:2010qj}
  The Heavy Flavor Averaging Group {\it et al.},
  arXiv:1010.1589 [hep-ex].


\bibitem{Abazov:2009wg}
   V.~M.~Abazov {\it et al.}  [D0 Collaboration],
  Phys.\ Rev.\  D {\bf 82}, 012003 (2010)
  [arXiv:0904.3907 [hep-ex]].

\bibitem{Grossman:2006ce}
  Y.~Grossman, Y.~Nir and G.~Raz,
  Phys.\ Rev.\ Lett.\  {\bf 97}, 151801 (2006)
  [arXiv:hep-ph/0605028].



\bibitem{Grinstein:2008qk}
  B.~Grinstein, K.~A.~Intriligator and I.~Z.~Rothstein,
  Phys.\ Lett.\  B {\bf 662}, 367 (2008)
  [arXiv:0801.1140 [hep-ph]].

\bibitem{Bona:2009tn}
  M.~Bona {\it et al.},
  arXiv:0906.0953 [hep-ph].

\bibitem{Gamiz:2009ku}
  E.~Gamiz, C.~T.~H.~Davies, G.~P.~Lepage, J.~Shigemitsu and M.~Wingate
                  [HPQCD Collaboration],
  Phys.\ Rev.\  D {\bf 80}, 014503 (2009)
  [arXiv:0902.1815 [hep-lat]].


\end{thebibliography}
\end{document}